\documentclass[%
 aip,
 amsmath,amssymb,
 reprint,%
]{revtex4-1}

\usepackage{amsmath}
\usepackage{amsthm}
\usepackage{amssymb}
\usepackage{graphicx}
\usepackage{dcolumn}
\usepackage{xcolor}
\usepackage{mathtools}
\usepackage{bm}

\usepackage[utf8]{inputenc}
\usepackage[T1]{fontenc}
\usepackage{mathptmx}
\usepackage{etoolbox}

\makeatletter
\def\@email#1#2{%
 \endgroup
 \patchcmd{\titleblock@produce}
  {\frontmatter@RRAPformat}
  {\frontmatter@RRAPformat{\produce@RRAP{*#1\href{mailto:#2}{#2}}}\frontmatter@RRAPformat}
  {}{}
}%

\usepackage{mathtools}
\usepackage{latexsym}
\usepackage{soul}
\usepackage{siunitx}
\usepackage{textgreek}
\usepackage{natbib} 
\usepackage{lineno}
\usepackage[utf8]{inputenc} 

\usepackage{lipsum}
\usepackage{cuted}
\usepackage{color}
\usepackage{xcolor}
\usepackage{comment}
\usepackage{float}
\usepackage{booktabs}
\usepackage{multirow}
\usepackage{tabularx}
\newcolumntype{L}{>{\raggedright\arraybackslash}X}

\usepackage{comment}
\usepackage{chngcntr}
\counterwithin*{equation}{part}
\counterwithin*{figure}{part}

\newcommand{\prt}[1]{%
  \stepcounter{part}\addtocounter{part}{-1}%
  \part*{#1}%
  \addcontentsline{toc}{part}{#1}%
}

\newcommand{\tgrad}{\boldsymbol{\nabla}}
\newcommand{\vgrad}{\boldsymbol{\nabla}}
\newcommand{\tr}{\operatorname{tr}}
\newcommand{\tpp}{\operatorname{T}}
\newcommand{\tp}[1]{#1^{\tpp}}
\newcommand{\tpgrad}{\tp{\boldsymbol{\nabla}}}
\newcommand{\lap}{\Delta}

\newcommand{\vdiver}{\operatorname{\mathbf{div}}\,}
\newcommand{\diver}{{\operatorname{div}}\,}
\newcommand{\vonkar}{von K\'arm\'an }

\makeatother
\begin{document}

\preprint{AIP/123-QED}

\title[]{Dynamics of pressurized ultra-thin membranes}
\author{Ali Sarafraz}
\affiliation{%
Department of Precision and Microsystems Engineering, Faculty of Mechanical, Maritime and Materials
Engineering, Delft University of Technology, 2628 CD, Delft, The Netherlands.
}
\email{a.sarafraz@tudelft.nl}

\author{Arthur Givois}%
\affiliation{%
Department of Precision and Microsystems Engineering, Faculty of Mechanical, Maritime and Materials
Engineering, Delft University of Technology, 2628 CD, Delft, The Netherlands.
}%
\affiliation{Université de technologie de Compiègne, Roberval (mechanics, energy and electricity) - Centre de recherche Royallieu - CS 60319 - 60203 Compiègne Cedex - France }

\author{Irek Rosłoń}
\affiliation{%
Department of Precision and Microsystems Engineering, Faculty of Mechanical, Maritime and Materials
Engineering, Delft University of Technology, 2628 CD, Delft, The Netherlands.
}%

\author{Hanqing Liu}
\affiliation{%
Department of Precision and Microsystems Engineering, Faculty of Mechanical, Maritime and Materials
Engineering, Delft University of Technology, 2628 CD, Delft, The Netherlands.
}%


\author{Hatem Brahmi}
\affiliation{%
ASML Netherlands B.V., 5504 DR, Veldhoven, The Netherlands.
}%

\author{Gerard Verbiest}
\affiliation{%
Department of Precision and Microsystems Engineering, Faculty of Mechanical, Maritime and Materials
Engineering, Delft University of Technology, 2628 CD, Delft, The Netherlands.
}%

\author{Peter G. Steeneken}
\affiliation{%
Department of Precision and Microsystems Engineering, Faculty of Mechanical, Maritime and Materials
Engineering, Delft University of Technology, 2628 CD, Delft, The Netherlands.
}%
\affiliation{%
Kavli Institute of Nanoscience, Faculty of Applied Sciences, Delft University of Technology, 2628 CJ, Delft, The
Netherlands
}%

\author{Farbod Alijani}
\affiliation{%
Department of Precision and Microsystems Engineering, Faculty of Mechanical, Maritime and Materials
Engineering, Delft University of Technology, 2628 CD, Delft, The Netherlands.
}%
\email{f.alijani@tudelft.nl}

\date{\today}

\begin{abstract}
The resonance frequency of ultra-thin layered nanomaterials changes nonlinearly with the tension induced by the pressure from the surrounding gas. Although the dynamics of pressurized nanomaterial membranes have been extensively explored, recent experimental observations show significant deviations from analytical predictions. Here, we present a multi-mode continuum approach to capture the nonlinear pressure-frequency response of pre-tensioned membranes undergoing large deflections. We validate the model using experiments conducted on polysilicon drums excited opto-thermally and subjected to pressure changes in the surrounding medium. We demonstrate that considering the effect of pressure on the membrane tension is not sufficient for determining the resonance frequencies. In fact, it is essential to also account for the change in the membrane's shape in the pressurized configuration, the mid-plane stretching, and the contributions of higher modes to the mode shapes. Finally, we show how the presented high-frequency mechanical characterization method can be used to determine Young's modulus of ultra-thin membranes.

\end{abstract}

\maketitle


Owing to their low bulk modulus and outstanding in-plane stiffness, sensors made of ultra-thin membranes and two-dimensional (2D) materials have recently gained interest for gas~\cite{Gupta2020, Varghese2015, Kim2017}, pressure~\cite{Smith2013, Koening2012, Dolleman2016, Dolleman20162, Atalaya2010}, and biosensing~\cite{Roslon2022} applications. Despite numerous experimental and theoretical studies~\cite{CastellanosGomez2015, Akinwande2017}, probing the elasticity of these membranes has remained challenging~\cite{Isacsson2016, Los2017}, making the development of new methods for their mechanical characterization of great importance~\cite{Davidovikj2017, Sajadi2017, Sajadi2019}.  



Atomic force microscopy (AFM) is the most widely used technique for extracting Young's modulus of 2D membranes, which is achieved by fitting their nonlinear force-deflection response ~\cite{Castellanos-Gomez2012, Lee2008, Cartamil-Bueno2017}. Pressurized blister test~\cite{Boddeti2013}, electrostatic deflection method~\cite{Nicholl2015, Verbiest2021}, and Duffing-type nonlinear response at large amplitude vibrations~\cite{Davidovikj2017, Sajadi2019} are other methods used to characterize thin membranes. Recently, it was also shown that the tension-induced shift in the resonance frequency of 2D membranes as a function of applied pressure could be viewed as a fast and contactless probe for characterizing elastic properties.~\cite{Lee2019}. In contrast to AFM which requires large indentations to enter the nonlinear regime and estimate Young's modulus~\cite{Komaragiri2005, Vella2017}, it was demonstrated that the geometrically nonlinear regime may be easily reached using this approach by sweeping the pressure across the membrane. 
However,  it was found that the fit of the pressure-frequency response, results in estimations of Young's modulus that are an order of magnitude lower than the well-accepted values in the literature~\cite{Dolleman2014, Lee2019}. Therefore, in order to use this method to characterize suspended ultra-thin materials that undergo large deflections, a more comprehensive model is required to describe the underlying physics.

\begin{figure}
 \centering
 \includegraphics[width=0.45\textwidth]{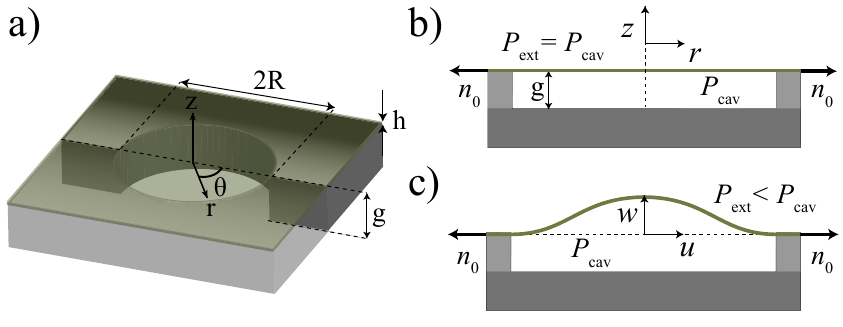}
 \caption{a) Schematic of the membrane and its cross-sections. b) Side-view of the undeformed membrane, and c) deformed configuration under transverse loading due to pressure difference alongside the membrane.}
 \label{fig:Schematic}
\end{figure}

 \begin{figure*}
 \centering
 \includegraphics[width=1\textwidth]{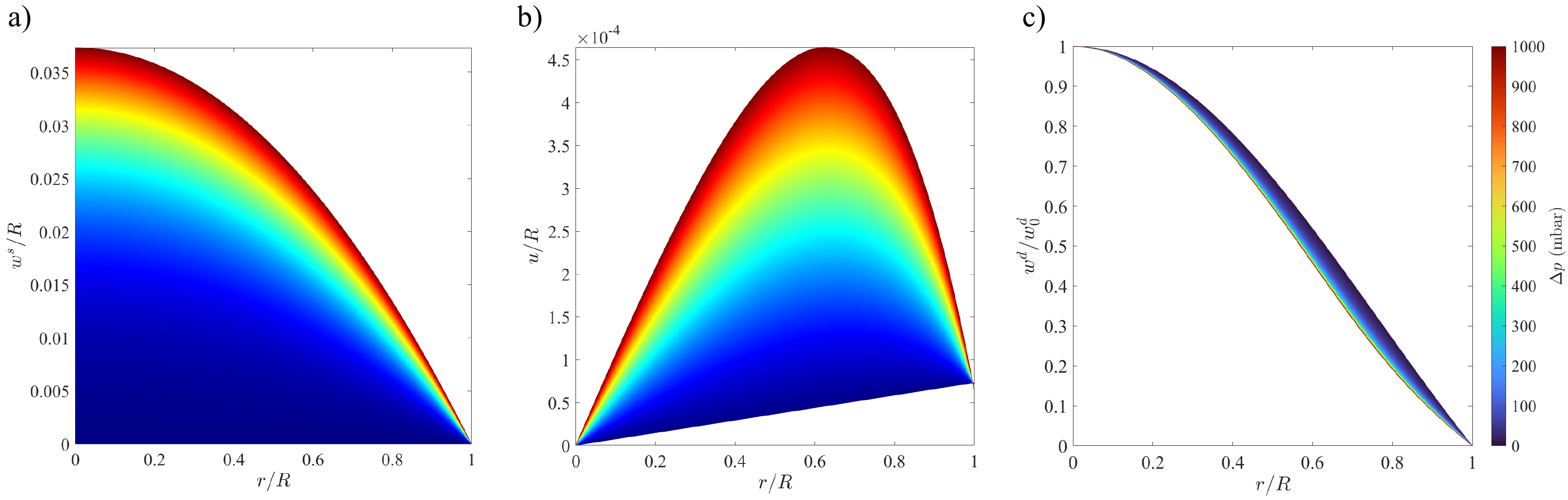}
 \caption{Variation of the static and dynamic displacement fields with pressure. a) Normalized static transverse displacement varying from a flat to a nearly parabolic configuration; 
 b) normalized in-plane displacement showing deviation from linear variation in the radial direction with increasing pressure; c) normalized mode shape as a function of external pressure obtained from the multi-mode model ($w_0^d$ denotes the dynamic transverse deflection at the center of the membrane). Here the proposed model is solved using modal coordinates $q_k$. Eqs. (S26-28) of the supplementary information were used to obtain in-plane ($u$) and out-of-plane displacements ($w^s$, and $w^d$).}
 \label{fig:ModeShape}
\end{figure*}

Here, we present a multi-modal continuum model that quantifies the dynamical behavior of ultra-thin membranes subjected to high pressures. Unlike models commonly used for characterizing pressurized membranes~\cite{Bunch2008, Lee2014}, the proposed model accounts for a change in the static equilibrium shape of the vibrating drum, mid-plane stretching, and in-plane and out-of-plane mode coupling. By comparing the model to the Finite Element Method (FEM) simulations of drums subjected to large deflections, we validate the analytic formulation and compare it to existing theoretical models in the literature. Finally, we acquire the nonlinear pressure-frequency response of polysilicon drums experimentally and demonstrate that the suggested multi-mode model can be utilized to extract Young's modulus from the high-frequency dynamic response of pressurized ultra-thin membranes.

We consider a thin circular drum with a diameter of $2R$ and thickness  $h \ll R$ clamped at the outer edge, as shown in Fig.\,\ref{fig:Schematic}(a). The drum is assumed to be made of a homogeneous and isotropic material of density $\rho$, Young's modulus $E$, and Poisson's ratio $\nu$. The drum is also assumed to be subjected to an axisymmetric tension ${n_0}$, and the pressure difference alongside the drum is $\Delta p$ (see Fig.~\ref{fig:Schematic}(b, c)). We only account for axisymmetric vibrations and suppose that the aspect ratio is sufficiently small (i.e., $h/R < 0.001$~\cite{Mansfield1989}) such that the effect of bending rigidity can be discarded, and the motion can be modeled using membrane theory~\cite{Thomas2008, amabili2008nonlinear}. Following the modelling presented in supplementary information S1, we use the Föppl–von Kármán equations for pre-stressed drums and follow typical discretization using the in-plane (radial) and out-of-plane (transverse) eigenfunctions of a flat circular membrane to obtain the following set of nonlinear differential equations: 

\begin{subequations}
\label{eq:FormUWProj1}
\begin{align}
 \ddot{q}_k + \omega_k^2 q_k +  \varepsilon   \sum_{i=1}^{N_w} \sum_{j=1}^{N_w} \sum_{l=1}^{N_w} \Gamma_{ijl}^k q_i q_j q_l  = Q_k, \label{eq:FormUWProj1T} \\
\Gamma_{ijl}^k = c_{ijl}^k + \sum_{p=1}^{N_u} \frac{b_{jl}^p a_{pi}^k}{\gamma_p^2}, 
\label{eq:FormUWProj1A} 
\end{align}
  \end{subequations}
for $k=1, 2, ..., N_w$, where $q_k$ denotes the normalized out-of-plane modal coordinates. 
$N_w$ and  $N_u$ are the number of transverse and radial generalized coordinates that will be retained in the analysis, respectively.  Moreover, $\omega_k$ and $\gamma_p$ denote the transverse and radial dimensionless eigenfrequencies. 
$(a_{pi}^k,b_{ij}^p, c_{ijl}^k)$ are the nonlinear coefficients of the model that stem from geometric nonlinearities and $Q_k$ is the projection of the applied pressure on the k$^{\text{th}}$ transverse mode of the drum. Finally, $\varepsilon$ is a dimensionless parameter that depends on the geometric and material properties. 
The modal coordinates $q_k$ are the unknowns in the Eq.~\eqref{eq:FormUWProj1}, whereas all other parameters are described in Section S1 of supplementary information.

The solution of the system of Eqs.~(\ref{eq:FormUWProj1}) is obtained by splitting displacements into static and dynamic components~\cite{Sajadi2017,Camier2009,Steeneken2021} (${q_k} = {q^{s}_k} + {q^{d}_k}$, with ${q^{s}_k}$, and ${q^{d}_k}$ the static and dynamic parts of the solution). This separation allows us to obtain the static deflection due to $Q_k$ and linearize Eqs.~(\ref{eq:FormUWProj1}) around the statically deflected configuration as follows:
\begin{subequations}
\label{eq:StatDynA}
\begin{align}
\omega _k^2q_k^s + \varepsilon \sum\limits_{i = 1}^{{N_w}} {\sum\limits_{j = 1}^{{N_w}} {\sum\limits_{l = 1}^{{N_w}} {\Gamma _{ijl}^kq_i^s} } } q_j^sq_l^s &= {Q_k},
\label{eq:StatDynA1}\\
\ddot q_k^d + \omega _k^2q_k^d + \varepsilon \sum\limits_{i = 1}^{{N_w}} {\sum\limits_{j = 1}^{{N_w}} {\sum\limits_{l = 1}^{{N_w}} {(2\Gamma _{jil}^k + \Gamma _{ijl}^k)q_j^sq_l^s} } q_i^d} & = 0. \label{eq:StatDynA2}
\end{align}
\end{subequations}

Here, solving for $q_k^s$ in  Eq.~\eqref{eq:StatDynA1} would yield the statically deformed configuration. 
Moreover, Eq.~\eqref{eq:StatDynA2} corresponds to the linearized dynamic equation around the deflected state. The shift in resonance frequencies of the drum due to applied pressure is obtained from Eq.~\eqref{eq:StatDynA2} using the procedure described in the supplementary information S2 (see Eqs.~(S17), (S20), and (S21)).




\begin{figure}
 \centering
 \includegraphics[width=0.49\textwidth]{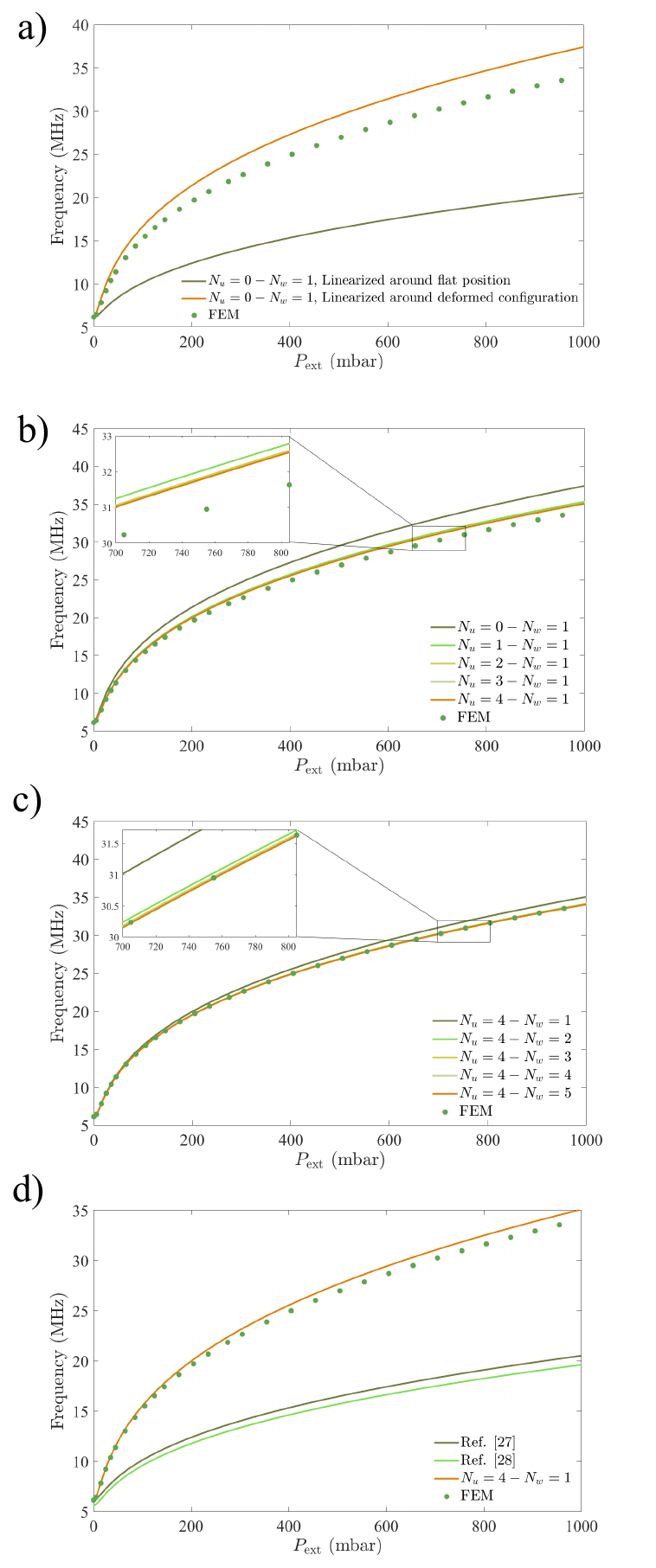}
 \caption{Comparison of the proposed model to the FEM result by a) linearizing the dynamic governing equation about the deformed static configuration rather than flat position; b) increasing the number of in-plane modes; c) addition of out-of-plane modes. d) Comparison to other models available in literature.}
 \label{fig:convergence}
\end{figure}


We simulate in Figs.~\ref{fig:ModeShape} the influence of pressure on the static and dynamic in-plane and out-of-plane displacements of a drum with $R=5~\SI{}{\mu\metre}$, $h=20~\SI{}{n\metre}$, $\rho=2300~ \SI{}{kg}.~\SI{}{\metre^{-3}}$, $E=160~\SI{}{GPa}$, $\nu=0.22$, $n_0=0.3~\SI{}{\newton/\metre}$. 
In Fig.~\ref{fig:ModeShape}(a) highlights that the original flat configuration changes significantly with increasing pressure. This 
shows that when estimating frequency shifts due to pressure, the dynamic governing equation must be linearized around the new bulged shape ($q_k^s\neq 0$) rather than the flat configuration ($q_k^s=0$).
The application of pressure causes a nonlinear in-plane displacement field at high pressures, as depicted in Fig.~\ref{fig:ModeShape}(b). In addition, Fig.~\ref{fig:ModeShape}~(c) illustrates the slight differences in dynamic mode shape that emerge due to statically deformed configurations. As we will discuss next, neglecting these effects could yield wrong estimation of the resonance frequencies of pressurized ultra-thin membranes.

\begin{figure*}
 \centering
 \includegraphics{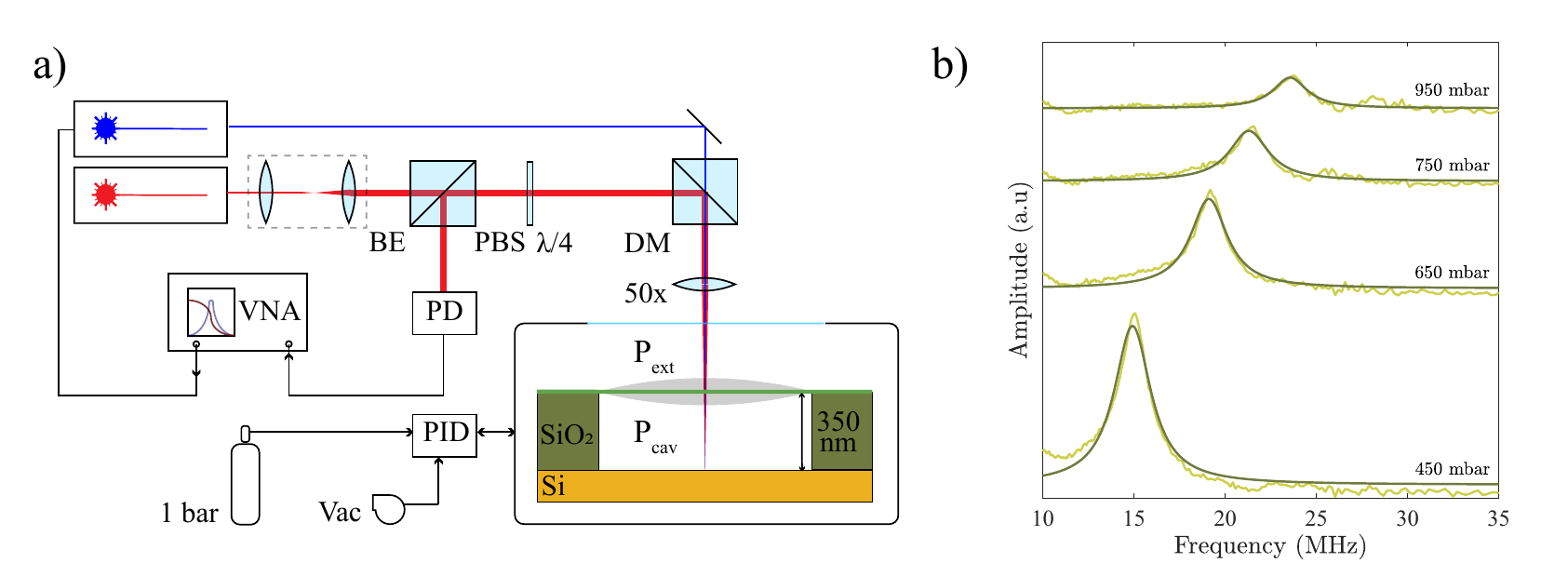}
 \caption{\textbf{The laser interferometry setup and obtained frequency spectra of polysilicon drums.} a) After passing through the polarized beam splitter (PBS) and the quarter-wave plate ($\lambda/4$), the red laser is combined with the blue laser and focused on the drum using a dichroic mirror (DM). The readout is performed by a high-frequency photodiode (PD), and the output is fed into the Vector Network Analyzer (VNA). The VNA modulates the blue laser that actuates the drum. A PID controller is used to regulate gas pressure inside the vacuum chamber. b) Frequency spectra (light green) are obtained from the VNA at different pressures. To determine the fundamental mode's resonance frequency, a damped harmonic oscillator model is fitted (dark green).}
 \label{fig:Experiment}
\end{figure*}

In order to highlight the influence of these effects on the estimation of the resonance frequencies, we benchmark in Fig.~\ref{fig:convergence}(a-c) the fundamental frequency derived from our model against FEM results and analytic solutions available in the literature. The simulations are performed for the same drum specifications as Fig.~\ref{fig:ModeShape}. In Fig.~\ref{fig:convergence}(a), we simulate the frequency shift 
where only one single out-of-plane mode is retained in the analysis ($N_u=0, N_w=1$). Frequency shifts are found, linearizing Eq.\eqref{eq:StatDynA2} around the undeformed (flat) as well as deformed configuration. The figure confirms our earlier prediction in Fig.~\ref{fig:ModeShape}(a) that linearizing around the deformed configuration substantially increases accuracy. When compared to FEM results, the model still has a maximum error of $9\%$ at 1 bar. 
To illustrate the effect of in-plane displacements on the predicted frequencies, we increase $N_u$ from 0 to 4 in Fig.~\ref{fig:convergence}(b) while retaining only one out-of-plane generalized coordinate ($N_w=1$) in the analysis.
It demonstrates that including more in-plane modes in the model still results in a more accurate model (the error relative to FEM simulations decreases to $4\%$ at 1 bar).
This confirms the important role of mid-plane stretching when tracing the fundamental resonance frequency of 2D drums as a function of pressure. In the limit of $N_u$=4 and $N_w$=1, it is  possible to simplify Eq.~\eqref{eq:StatDynA2} and devise the following set of simple analytic expressions for the pressure-displacement relation and the frequency of the fundamental mode of the drum in the pressurized state (see supplementary information S2 for the details):
\begin{subequations}
\label{eq:GenEqA}
\begin{align}
\Delta p = 3.61\frac{{{n_0}}}{{{R^2}}}\tilde q_1^s + 0.737\varpi (\nu )\frac{{Eh}}{{{R^4}}}{\left( {\tilde q_1^s} \right)^3},
\label{eq:GenEqA1}\\
f = \frac{{2.4048}}{{2\pi R}}\sqrt {\frac{{{n_0}}}{{\rho h}} + 0.612\varpi (\nu )\frac{E}{{\rho {R^2}}}{{\left( {\tilde q_1^s} \right)}^2}}. \label{eq:GenEqA2}
\end{align}
\end{subequations}
where $\varpi (\nu )$ is a function depending on the Poisson's ratio (see  Fig.~(S1) of the supplementary information), and $\tilde{q}_1^s = hq_1^s$ is the dimensional static deflection at the center of the membrane 
and $f$ is the new fundamental frequency.

Although the response from this model is already close to the FEM simulations, Fig.~\ref{fig:convergence}(c) shows that the accuracy is even further improved by including more transverse degrees-of-freedom in the basis functions by increasing $N_w$ in Eqs.~\eqref{eq:StatDynA}.
This finally leads to a negligible difference between simulations based on the present multi-mode model and FEM results, which highlights a slight contribution of higher order out-of-plane modes to the estimated resonance frequency. 

We finally compare the result from our single-transverse-mode model (Eq.~\eqref{eq:GenEqA}) to analytic models available in the literature, which are often used for the dynamic characterization of pressurized drums~\cite{Bunch2008, Lee2014} (see Fig.~\ref{fig:convergence}(d)). The substantial differences observed between our model and those of Ref.~\cite{Bunch2008, Lee2014} stem from the simplifying assumptions that include the absence of mid-plane stretching and linearization about flat configuration, which are both lifted in our analysis (see supplementary information S2 for the detailed comparison between the present multi-mode model and the analytic solutions available in the literature).

\begin{figure*}
 \centering
 \includegraphics[width=0.95\textwidth]{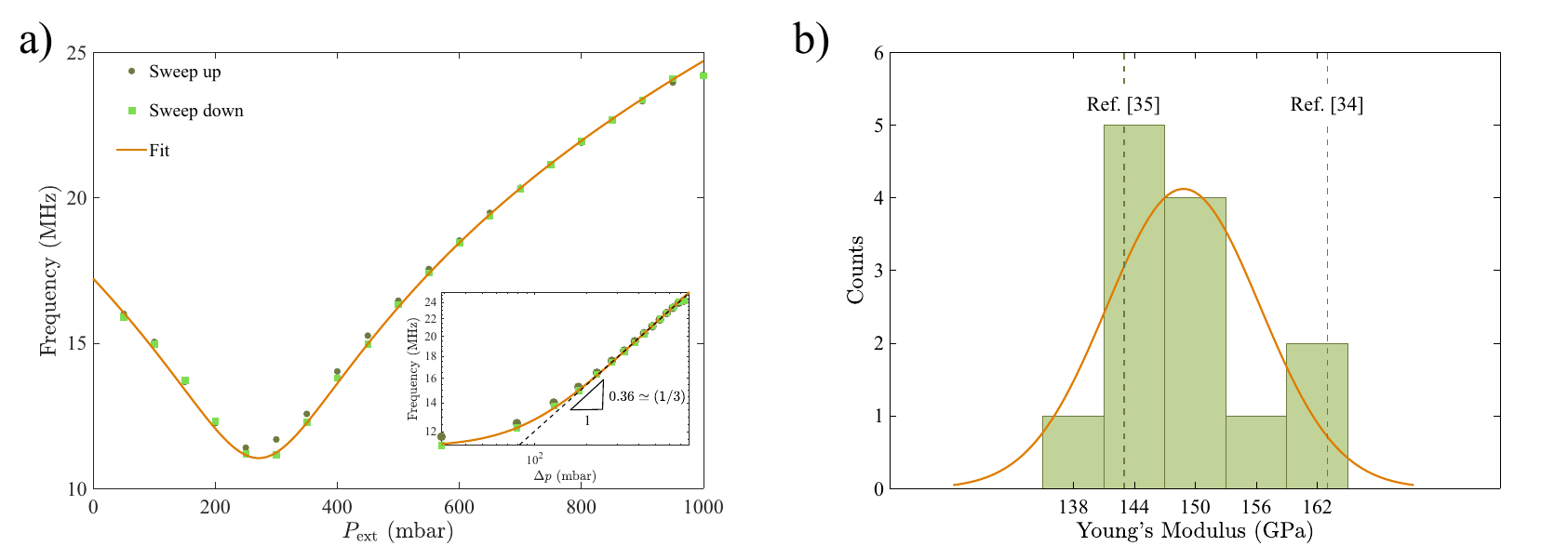}
 \caption{\textbf{Experimental result obtained for polysilicon drums.} a) Pressure-frequency of a drum in both sweeping up and down of pressure, fitted with our model with E = 155 GPa. b) Histogram of the Young’s moduli obtained through fitting the experimental data with the average value of 148 GPa, and standard deviation of 7 GPa.}
 \label{fig:Results}
\end{figure*}

To verify the applicability of the proposed formulation for the dynamic characterization of ultra-thin drums, we measure polysilicon membranes transferred over SiO2 substrate cavities, which are created using Reactive ion etching with a depth of 350 nm. The 10 \textmu m-diameter drums are placed in a vacuum chamber with variable pressure ranging from 50 to 1000 mbar. We employ a modulated blue laser diode ($\lambda = 405$ nm) to push the polysilicon drums into resonance (Fig.~\ref{fig:Experiment}a).
The suspended drum modulates the intensity of the reflected red laser ($\lambda = 633$ nm), which is collected at the photodiode and analyzed by a Vector Network Analyser (VNA). Our setup includes a PID controller that monitors chamber pressure using a vacuum pump and gas supply (nitrogen).
Fig.~\ref{fig:Experiment}(b) depicts the frequency spectra of a polysilicon drum at different pressures $\Delta P = {P_{{\rm{ext}}}} - {P_{{\rm{cav}}}}$, where $P_{{\rm{cav}}}$ and $P_{{\rm{ext}}}$ are the pressure outside and inside the cavity, respectively (see Fig.~\ref{fig:Schematic}). As a result of increased pressure, we observe an apparent tension-induced increase in the drum's resonance frequency. To obtain the pressure-frequency response, we sweep the pressure from $50$ to $1000$ mbar and fit the resonance peaks using the linear harmonic oscillator model to get resonance frequencies.
Fig.~\ref{fig:Results}(a) shows typical pressure-frequency measurement data and the model fit (E = 155 GPa, $n_0=1.6~\SI{}{\newton/\metre}$). In contrast to Fig.~\ref{fig:convergence}, we see a minimum in the pressure-frequency response at $P_{{\rm{cav}}}= 248$, which corresponds to a flat configuration. In this configuration, the drum's fundamental frequency is determined solely by pre-tension. With increasing pressure, the drum deforms statically, and the resonance frequency varies nonlinearly to $f \propto \sqrt[6]{E}\Delta {p^{1/3}}$ (see supplementary information S2). Using pressure measurements, one may determine the drum's Young's modulus.

To demonstrate the versatility of the method for characterizing Young's modulus of ultra-thin membranes, we repeated the same measurements on 13 drums that were adequately sealed (or the leak rate was low enough relative to the experiment's data collection speed) and did not exhibit hysteresis in the pressure-frequency response when the pressure was swept up and down. Fig.~\ref{fig:Results}(b) depicts the retrieved Young's moduli histogram. We see that the obtained average value $E = 148 \pm 7$\,GPa from our technique is comparable to the literature-reported uni-axial stretching test findings for thin polysilicon beams~\cite{Hemker2007, Corigliano2004}, further validating the accuracy of our model. In addition, as the second fitting value, we determined $n_0 = 1.1 \pm 0.5\,\SI{}{{\newton/\metre}}$.

Employing the pressure-frequency response as a characterization method has several advantages over commonly used characterization techniques. First, AFM requires considerable deflections to estimate Young's modulus, i.e., the force-deflection curve should reach a cubic regime~\cite{Komaragiri2005, Wan2003, Vella2017}($F \propto \delta ^3$; $F$: tip force, $\delta$: membrane's center deflection). In contrast, pressurization using the surrounding gas requires a lower force for determining Young's modulus and distributes the load over the membrane, minimizing the chance of membrane failure. The method also reaches the nonlinear regime suitable for estimating Young's modulus with substantially less deflections and stresses, thus making it a more practical method for characterizing brittle nanomaterials such as polysilicon (see supplementary information S3). Furthermore, unlike nonlinear dynamic characterization, which uses large driving signals to push the membrane into a nonlinear Duffing regime and requires proper calibration of the vibration amplitude~\cite{Davidovikj2017}, this method is independent of the vibration amplitude and only measures the resonance frequency as a function of the applied pressure. 

Several points should be kept in mind when employing the proposed method for characterization. First, in an optical detection scheme, the cavity depth has to be optimized so that the photodiode voltage is still linearly related to the motion at high amplitudes~\cite{Barton2012, Keskekler2021}. Moreover, gas leakage through the clamped boundaries and poor adhesion forces between the drum and substrate should be avoided to ensure negligible leakage rates and to prevent significant variations of the internal cavity pressure during the measurement~\cite{Lee2019, Bunch2008_2, Sun2020, Manzanares-Negro2020, Lee2022}.




Finally, when ${{\rm{P}}_{cav}}{\rm{ > 0}}$, the cavity pressure change with the membrane's static deflection, and the squeeze film effect~\cite{Pratap2007} must be considered while analyzing the dynamics of pressurized ultra-thin drums. Although both effects contribute significantly to the tension induced in the membrane, their cumulative effect on the drum's frequency shift near 1 bar and in the high-pressure regime is negligible for our polysilicon drums, keeping the extracted value of Young's modulus unchanged. We also note that neglecting the squeeze-film effect and the bending rigidity of a relatively thick membrane would result in an overestimation of pre-tension; therefore, accounting for these two effects will enhance the accuracy of the anticipated pre-tension value (see supplementary information S4 for more details).

\section*{Conclusion}
In summary, attempts to use frequency shifts of pressurized drums for characterization purposes failed due to a lack of appropriate mathematical models. A multi-mode continuum model for predicting the dynamics of ultra-thin pre-tensioned drums subjected to high pressures is presented here. By maintaining in-plane and out-of-plane membrane deformations, we investigate the pressure-dependent resonance frequency and mode shapes and illustrate the crucial importance of mid-plane stretching in the high-pressure regime. We highlight the discrepancies by comparing the model's accuracy to previously reported analytical models and FEM simulations. By fitting the resonance frequency measurements of a series of polysilicon drums as a function of the applied pressure as determined by laser interferometry, we demonstrate the model's validity and applicability for determining Young's modulus of ultra-thin nanomaterial membranes using on-resonance high-frequency characterization. Since pressurization results in a distributed stress field compared to indentation, this method could be utilized as an effective tool for determining the Young's modulus of brittle nanomaterials.


\section*{ACKNOWLEDGMENTS}

The authors would like to thank Dr. Martin Lee for fruitful discussions about sealing problems in our experiments. We thank Dr. Robin Dolleman and Dr. Dejan Davidovikj for their earlier investigations into this topic. This project has received funding from European Union’s Horizon 2020 research and innovation programme under Grant Agreement Nos. 802093 (ERC starting grant ENIGMA), 785219, and 881603 (Graphene Flagship).

\section*{AUTHOR DECLARATIONS}

The authors have no conflicts to disclose.

\section*{Data availability}

The data that support the findings of this study are available from the corresponding author
upon reasonable request.

\newpage

\renewcommand{\thepage}{S\arabic{page}} 
\renewcommand{\thesection}{S\arabic{section}}  
\renewcommand{\thetable}{S\arabic{table}}  
\renewcommand{\thefigure}{S\arabic{figure}}
\renewcommand{\theequation}{S\arabic{equation}}

\prt{Supplementary information}
\section{Model derivation}
\label{Sec:NonlinearCoefs}
As stated in the manuscript, this study focuses on a thin circular membrane with  diameter $2R$ and  thickness $h \ll R$, made of a homogeneous and isotropic material with  density $\rho$, Young's modulus $E$, and Poisson's ratio $\nu$. This thin structure is assumed to be subjected to an axisymmetric tension ${n_0}$. The equations of motion for a membrane subjected to large transverse displacement $w$, moderate rotations, are the dynamic equivalents of the \vonkar equations. In this context, the governing equations of a pre-stressed membrane with negligible bending rigidity are \cite{Reddy2003, amabili2008nonlinear}:
\begin{subequations}
\label{eq:sysNLPlak}
\begin{align}
\rho h \ddot{w} -n_0 \nabla^2 w - \diver(\bm{N}\vgrad w) & = {\Delta p},  \\
\rho h \ddot{\bm{u}} - \vdiver\bm{N} & = 0. 
\end{align}
\end{subequations} 
where:
\begin{equation}
\label{eq:Neps}
\begin{array}{r}
\bm{N}=[Eh/(1- \nu^2)]\left[(1-\nu)\bm{\epsilon}+\nu\tr\bm{\epsilon}\bm{1}\right], \\
\bm{\epsilon}=\frac{1}{2}\left(\tgrad\bm{u}+\tpgrad\bm{u} + \tgrad w\otimes\tgrad w\right).
\end{array}
\end{equation}

Here, ${\Delta p}$ represents pressure difference, ${P_{cav}-P_{ext}}$, alongside the membrane (see Fig.1 (b) of the main article), and $\bm{u}=[  u ;  v ]$ is the axial displacements. Moreover, $\nabla^2$, $\tgrad w$ and $\tgrad \bm{u}$ denote the Laplacian of scalar field $w$, the vector gradient of scalar field $w$ and the tensor gradient of the vector field $\bm{u}$, respectively. $\text{div} \bm{u}$ and $\vdiver \bm{N}$ are the scalar and vector divergences of vector field $\bm{u}$ and tensor field $\bm{N}$. Finally, $\tgrad w\otimes\tgrad w$ corresponds to the tensor product between vectors $\tgrad w$ and $\tgrad w$. Overdot also indicates derivation with respect to time. By axisymmetric vibrations the following set of equations can be obtained in cylindrical coordinates and in terms of radial ($u$) and transverse ($w$) displacements~\cite{Reddy2006}:\begin{subequations}
\label{eq:eqNLPlaqUWAxi}
\begin{align}
\begin{array}{l}
\rho h\ddot w - n_0 \lap w - \left( {{{Eh} \mathord{\left/
 {\vphantom {{Eh} {\left( {1 - {\nu ^2}} \right)}}} \right.
 \kern-\nulldelimiterspace} {\left( {1 - {\nu ^2}} \right)}}} \right)[{u_{,rr}}{w_{,r}} + {u_{,r}}{w_{,rr}} \\
\,\,\,\,\,\,\,\,\,\,\,\,\,\,\,\,\,\,\,\,\,\,\,+ (1 + \nu )\frac{{{u_{,r}}}}{r}{w_{,r}} + \nu \frac{u}{r}{w_{,rr}} + \frac{{{{({w_{,r}})}^3}}}{{2r}} + \frac{3}{2}{({w_{,r}})^2}{w_{,rr}}] = {{\Delta p}},
\end{array}
\label{eq:eqNLPlaqUW1Axi}\\
\left( {{{Eh} \mathord{\left/
 {\vphantom {{Eh} {\left( {1 - {\nu ^2}} \right)}}} \right.
 \kern-\nulldelimiterspace} {\left( {1 - {\nu ^2}} \right)}}} \right)[{u_{,rr}} + \frac{{{u_{,r}}}}{r} - \frac{u}{{{r^2}}} + \frac{{1 - \nu }}{2}\frac{{{{({w_{,r}})}^2}}}{r} + {w_{,r}}{w_{,rr}}] = 0, \label{eq:eqNLPlaqUW2Axi}
\end{align}
\end{subequations}where $_{,r}$ and $_{,rr}$ denote the first and second derivatives with respect to $r$. To write Eqs. (\ref{eq:eqNLPlaqUWAxi}a, b) in dimensionless form, we introduce the following  set of dimensionless variables:
\begin{equation} \begin{array}{l}
\bar w = \frac{w}{h},\quad \bar u = \frac{R}{{{h^2}}}u,\quad \bar r = \frac{r}{R},\\
\bar t = \frac{1}{R}\sqrt {\frac{{{n_0}}}{{\rho h}}} t, \quad \overline {\Delta p}  = \frac{{{R^2}}}{{{n_0}h}}{{\Delta p}},\quad \varepsilon   = \frac{1}{{1 - {\nu ^2}}}\frac{{Y{h^3}}}{{{n_0}{R^2}}},
\end{array} \label{eq:AdimPlak}
\end{equation}
This would yield the following  non-dimensional form of Eq.~\eqref{eq:eqNLPlaqUWAxi}:
\begin{subequations}
\label{eq:NLsysPlate}
\begin{align}
\begin{array}{r}
\ddot{\bar{w}} - \Delta \bar w - \varepsilon \left[ {{{\bar u}_{,rr}}{{\bar w}_{,r}} + \left( {1 + \nu } \right)\frac{{{{\bar u}_{,r}}}}{r}{{\bar w}_{,r}} + \nu \frac{{\bar u}}{r}{{\bar w}_{,rr}} + \frac{{{{\left( {{{\bar w}_{,r}}} \right)}^3}}}{{2r}}} \right.\\
\left. { + \frac{3}{2}{{\left( {{{\bar w}_{,r}}} \right)}^2}{{\bar w}_{,rr}}} \right] = \overline {\Delta p} 
\end{array} ,
\label{eq:eqNLPlaqUW1AxiAdim}\\
    {\bar u_{,rr}} + \frac{{{{\bar u}_{,r}}}}{r} - \frac{{\bar u}}{{{r^2}}} + \frac{{1 - \nu }}{2}\frac{{{{\left( {{{\bar w}_{,r}}} \right)}^2}}}{r} + {\bar w_{,r}}{\bar w_{,rr}} = 0. \label{eq:eqNLPlaqUW2AxiAdim}
\end{align}
\end{subequations}
In order to solve Eqs. (\ref{eq:NLsysPlate}a, b), we first expand the transverse displacement $w$ and radial displacement $u$ as follows:\begin{subequations}
\label{eq:DecomposBMMembr}
\begin{align}
\bar w(r,t) = \sum\limits_{k = 1}^{{N_w}} {{\Phi _k}\left( r \right){q_k}(t){\mkern 1mu} {\mkern 1mu} }, \label{eq:DecomposBMMembrA}\\
\bar u(r,t) = \sum\limits_{p = 1}^{{N_u}} {{\Psi _p}\left( r \right){\eta _p}(t)} \label{eq:DecomposBMMembrB},
\end{align}
\end{subequations}
where $q_k$ and $ \eta_p$ denote the unknown modal coordinates, and $N_w$ and  $N_u$ are the number of generalized coordinates that will be retained in the analysis. Moreover, $\Phi _k$ and $\Psi _p$ are mode shapes associated with the transverse and radial displacements, respectively. These modes are chosen such that they satisfy the following eigenvalue problems that stem from the linear counterpart of Eqs.~(\ref{eq:NLsysPlate}):
\begin{equation}
\label{eq:Membrep}
\lap \Phi_k + \omega_k^2 \Phi_k= 0, \qquad
\Psi_{,rr} + \frac{\Psi_{,r}}{r} - \frac{\Psi}{r^2} + \gamma_p^2 \Psi_p = 0
\end{equation}
with $\omega_k$ and $\gamma_p$ being the transverse and in-plane dimensionless eigenfrequencies, respectively. We note that Eqs.~\eqref{eq:Membrep} are Bessel equations, and their solution can be expressed as:
\begin{equation}
\label{eq:ModeShapes}
    \Phi_k(r) = \kappa_k J_0 (\omega_k r), \qquad  \Psi_p(r) = \lambda_p J_1 (\gamma_p r).
\end{equation}
These modes have orthogonality properties and are normalized with respect to the modal mass, by fixing the values of $\kappa_k$, $\lambda_p$ introduced in Eqs.~ \eqref{eq:ModeShapes} and \eqref{eq:Membrep}. They are chosen such that:
\begin{equation}\label{eq:orth}
\iint_S \Phi_{i} \Phi_{j} \text{d} S = \delta_{ij} \qquad \text{and} \qquad \iint_S \Psi_{p}\Psi_{l} \text{d} S = \delta_{pl}.
\end{equation}
where $S$ is the domain of integration. By applying Galerkin technique, namely inserting Eqs.~\eqref{eq:ModeShapes}, and Eqs.~\eqref{eq:DecomposBMMembr} in Eqs.~\eqref{eq:NLsysPlate}, multiplying Eq.~\eqref{eq:eqNLPlaqUW1AxiAdim} by $\Phi_k$ and \eqref{eq:eqNLPlaqUW2AxiAdim} by $\Psi_p$, and then integrating over the entire domain, the nonlinear partial differential Eqs.~\eqref{eq:eqNLPlaqUW1AxiAdim} and \eqref{eq:eqNLPlaqUW2AxiAdim} reduce to:
\begin{subequations}
\label{eq:FormUWProj}
\begin{align}
 \ddot{q}_k + \omega_k^2 q_k + \varepsilon \sum_{p=1}^{N_u} \sum_{i=1}^{N_w} a_{pi}^k \eta_p q_i + \varepsilon  \sum_{i=1}^{N_w} \sum_{j=1}^{N_w} \sum_{l=1}^{N_w} c_{ijl}^k q_i q_j q_l   & = Q_k \label{eq:FormUWProjT} \\
\gamma_p^2 \eta_p - \sum_{i=1}^{N_w} \sum_{j=1}^{N_w} b_{ij}^p q_i q_j & = 0     \label{eq:FormUWProjA} 
\end{align}
  \end{subequations}
  where ${Q_k} = \overline {\Delta p} {\left[ {\int_0^1 {\Phi _k^2r{\rm{d}}r} } \right]^{ - 1}}\left[ {\int_0^1 {{\Phi _k}r{\rm{d}}r} } \right]$ is the projection of ${\overline {\Delta p}}$ on the k-th mode of vibration, and the nonlinear modal coefficients $a_{pi}^k,b_{ij}^p$, and $c_{ijl}^k$  are:

\begin{widetext}
\begin{subequations}
\begin{flalign}
 a_{pi}^k = & - \bigg[ \int_0^1 {\Phi _k^2r{\rm{d}}r} \bigg ]^{-1} {\int_0^1 {{\Phi _k}} \bigg({\Psi _{p,rr}}{\Phi _{i,r}} + {\Psi _{p,r}}{\Phi _{i,rr}} + (1 + \nu )\frac{{{\Psi _{p,r}}{\Phi _{i,r}}}}{r} + \nu \frac{{{\Psi _p}{\Phi _{i,rr}}}}{r}\bigg)r{\rm{d}}r},\\
b_{ij}^p = & + \frac{1}{2} \bigg[ \int_0^1 {\Psi _p^2r{\rm{d}}r} \bigg ]^{-1} {\int_0^1 {{\Psi _p}} \bigg(\frac{{1 - \nu }}{r}{\Phi _{i,r}}{\Phi _{j,r}} + {{({\Phi _{i,r}}{\Phi _{j,r}})}_{,r}} \bigg ) r{\rm{d}}r} ,\\
c_{ijl}^k = & - \frac{1}{2} \bigg[ \int_0^1 {\Phi _k^2r{\rm{d}}r} \bigg ]^{-1}  \int_0^1 {{\Phi _k}} \bigg({\Phi _{i,r}}{\Phi _{j,r}}{\Phi _{l,r}} + \frac{{{{({\Phi _{i,r}}{\Phi _{j,r}}{\Phi _{l,r}})}_{,r}}}}{r} \bigg ) r{\rm{d}}r. 
\end{flalign}
\end{subequations}
\end{widetext}

In order to find $\eta_p$ in terms of $q_k$, we rewrite Eq.~\eqref{eq:FormUWProjA}:
\begin{equation} 
 {\eta _p} = \frac{1}{\gamma _p^2} \sum\limits_{i = 1}^{{N_w}} {\sum\limits_{j = 1}^{{N_w}} {b_{ij}^p{q_i}{q_j}}}.   
\label{eq:ContInPlane}
\end{equation}
Inserting Eq.~\eqref{eq:ContInPlane} in Eq.~\eqref{eq:FormUWProjT} \cite{Davidovikj2017,Givois2019} leads to:
  \begin{equation} 
\ddot{q}_k + \omega_k^2 q_k +  \varepsilon   \sum_{i=1}^{N_w} \sum_{j=1}^{N_w} \sum_{l=1}^{N_w} \Gamma_{ijl}^k q_i q_j q_l  = Q_k,  
\label{eq:FormCondProj}
\end{equation}
with
\begin{equation}
\Gamma_{ijl}^k = c_{ijl}^k + \sum_{p=1}^{N_u} \frac{b_{jl}^p a_{pi}^k}{\gamma_p^2}.\label{eq:CondensCoeffPoutre}
\end{equation}
Eq.~(\ref{eq:FormCondProj}) is a reduced-order model, which includes mid-plane stretching and higher-mode interactions. 

\section{Solution of the governing equations}
\label{Sec:Solution}


To consider the dynamic oscillations around the static shape induced by pressure, we first split the contribution of static (${q_k^s}$) and dynamic (${ q_k^d}$) parts as follows~\cite{Camier2009, Sajadi2017}:
\begin{equation}
{q_k} = {q_k^s} + { q_k^d}. \label{eq:CondensCoeffPoutre}
\end{equation}

This separation (Eq. (\ref{eq:CondensCoeffPoutre})) allows us to separate Eq. (\ref{eq:FormCondProj}) to a static and dynamic part as well. We can obtain the static deflection due to $Q_k$, and linearize the equation of motion around the statically deflected configuration that shall be used for obtaining the natural frequencies of the drum in the deformed state and as a function of the applied pressure, as follows:
\begin{subequations}
\label{eq:StatDyn}
\begin{align}
\omega _k^2q_k^s + \varepsilon \sum\limits_{i = 1}^{{N_w}} {\sum\limits_{j = 1}^{{N_w}} {\sum\limits_{l = 1}^{{N_w}} {\Gamma _{ijl}^kq_i^s} } } q_j^sq_l^s = {Q_k},
\label{eq:StatDyn1}\\
\ddot q_k^d + \omega _k^2q_k^d + \varepsilon \sum\limits_{i = 1}^{{N_w}} {\alpha _i^kq_i^d} = 0, \label{eq:StatDyn2}
\end{align}
\end{subequations}
where
\begin{equation} 
{\alpha _i^k = \sum\limits_{j = 1}^{{N_w}} {\sum\limits_{l = 1}^{{N_w}} {(2\Gamma _{jil}^k + \Gamma _{ijl}^k)q_j^sq_l^s} } .}
\label{eq:AlphaBeta}
\end{equation}

In order to solve Eq.~(\ref{eq:StatDyn1}) for $q_k^s$, we use the software package MANLAB~\cite{Guillot2019}. This employs the Asymptotic Numerical Method (ANM)~\cite{Karkar2012, Cochelin2007, Cochelin1994}, a continuation algorithm based on high-order Taylor series expansions, to solve nonlinear systems of equations. The ANM continuation method allows for solving nonlinear algebraic equations written in the form:\begin{equation} 
\bm{F} (\bm{U},\lambda ) = \bm{0},
\label{eq:continuation}
\end{equation}
where $\bm{F}$ denotes a set of nonlinear algebraic equations, $\bm{U}=[q_1^s,q_2^s, \ldots ,q_{{N_w}}^s]$ is a vector of unknown coefficients, and $\lambda$ is the continuation parameter. The method also requires the system of equations to be written with quadratic nonlinearities \cite{Guillot2019, Karkar2012, Cochelin2007}. We can proceed with the continuation technique 
by recasting the static part of Eq.~(\ref{eq:StatDyn1}) to:
\begin{subequations}
\label{eq:FormUWProjStat}
\begin{align}
\omega _k^2q_k^s + \varepsilon \sum\limits_{i = 1}^{{N_w}} {\sum\limits_{j = 1}^{{N_w}} {\sum\limits_{l = 1}^{{N_w}} {\Gamma _{ijl}^k} } } {S_{jl}}q_i^s - \lambda {Q_k} & = 0,\label{eq:FormUWProjTStat} \\
{S_{jl}} - q_j^sq_l^s & = 0.     \label{eq:FormUWProjStatA} 
\end{align}
\end{subequations}
By solving Eq.~(\ref{eq:FormUWProjStat}) for $q_k^s$, one can find the shifted frequency due to pressure.
We note that the fundamental frequency of the pressurized membrane ${\Omega _u}$ are obtained by considering the pre-factors of $q_k^d$ in Eq.~(\ref{eq:StatDyn2}) which is denoted with the matrix $\bm{A}$, defined as:
\begin{equation} 
{A_{uv}} = \omega _u^2{\delta _{uv}} + \varepsilon \alpha _u^v.
\label{eq:MatrixA}
\end{equation}
where $\delta_{uv}$ is the kronecker delta. Diagonalizing the matrix $\bm{A}$ leads to the modified frequencies and mode shapes of the bulged membrane with an initial bulged shape:
\begin{equation} 
{\left[ {\Omega _u^2{\delta _{uv}}} \right]_{u,v \in \left[ {1,{N_w}} \right]}} = {\bm{P}^{ - 1}}\bm{A}\bm{P}
\label{eq:ModifiedFreq}
\end{equation}
with $\bm{P}$ the eigenvectors associated with the frequencies ${\Omega _u}$ for the pressurized membrane . 

\begin{figure}[tb]
 \centering
 \includegraphics[width=0.5\textwidth]{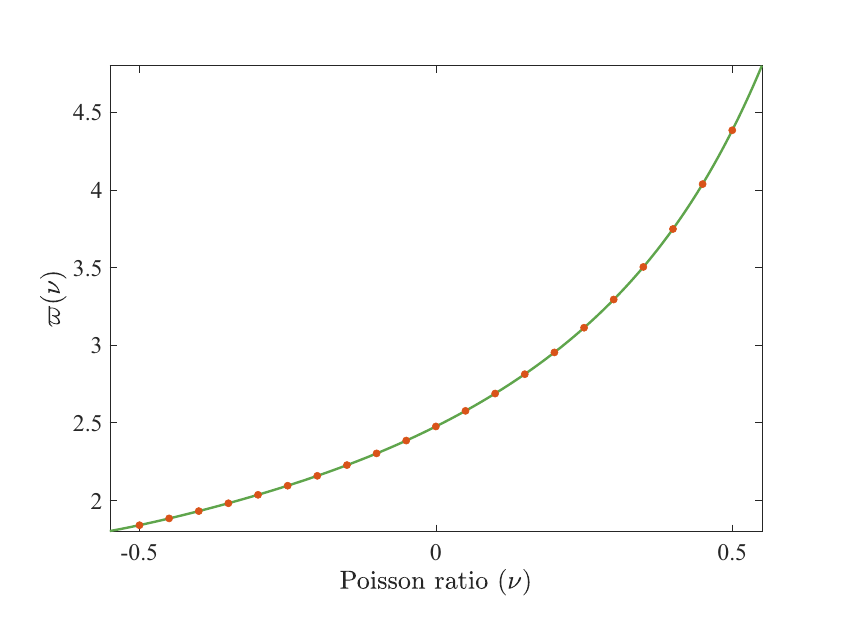}
 \caption{\textbf{Function $\varpi (\nu)$ used in single-transverse-mode analysis} (see Eq. (\ref{eq:GenEq})). The function is evaluated for finite numbers of Poisson ratios and interpolated with the cubic spline method.}
 \label{fig:varpi}
\end{figure}
Considering only $N_w=1$ and ${N_u}=4$, and converting equations to dimensional form using Eq.~\eqref{eq:AdimPlak} yields the following equations as a simplified version of Eqs.\eqref{eq:StatDyn}:
\begin{subequations}
\label{eq:duffing}
\begin{align}
\frac{{5.7831{n_0}}}{{\rho h{R^2}}}\tilde q_1^s + 1.181\frac{{E\varpi \left( \nu  \right)}}{{\rho {R^4}}}{\left( {\tilde q_1^s} \right)^3} = \frac{{1.6019}}{{\rho h}}\Delta p,\\
\ddot{\tilde{q}}_1^d + \frac{{5.7831{n_0}}}{{\rho h{R^2}}}\tilde q_1^d + 3.543\frac{{E\varpi \left( \nu  \right)}}{{\rho {R^4}}}{\left( {\tilde q_1^s} \right)^2}\tilde q_1^d = 0, \label{eq:duffing2}
\end{align}
\end{subequations}
where $\varpi (\nu )$ is plotted in Fig.~(\ref{fig:varpi}). Moreover, $\tilde q_1^s = hq_1^s$ denotes dimensional static deflection at the center of the membrane and $\tilde q_1^d = hq_1^d$ is the dimensional dynamic deflection at the center of the membrane. For this simplified scenario, the nonlinear pressure-deflection and pressure-frequency relationships are derived analytically by linearizing Eq.\eqref{eq:duffing} around the deflected configuration as follows:
\begin{subequations}
\label{eq:GenEq}
\begin{align}
\Delta p = A\frac{{{n_0}}}{{{R^2}}}\tilde q_1^s + B\frac{{Eh}}{{{R^4}}}{\left( {\tilde q_1^s} \right)^3},
\label{eq:GenEq1}\\
f = \frac{{2.4048}}{{2\pi R}}\sqrt {\frac{{{n_0}}}{{\rho h}} + C\frac{E}{{\rho {R^2}}}{{\left( {\tilde q_1^s} \right)}^2}}. \label{eq:GenEq2}
\end{align}
\end{subequations}
A simple initial estimation for fitting the pressure-frequency response of nanodrums is provided by Eq.~(\ref{eq:GenEq}) and Fig.~(\ref{fig:varpi}). While Eq.~(\ref{eq:GenEq2}) obtains the frequency of the nanodrum about the deflected configuration, Eq.~(\ref{eq:GenEq1}) defines the equilibrium position in the pressurized state.
In Table~(\ref{table:parameters}) we benchmark this new formulation against the analytic solutions available in the literature for pressurized nanodrums 
\cite{Bunch2008, Lee2014}. We note that the deviations that are observed in the parameter values of the models is the outcome of neglecting mid-plane stretching and lineariziation about the flat configuration. 

 \begin{figure}[t]
 \centering
 \includegraphics[width=0.5\textwidth]{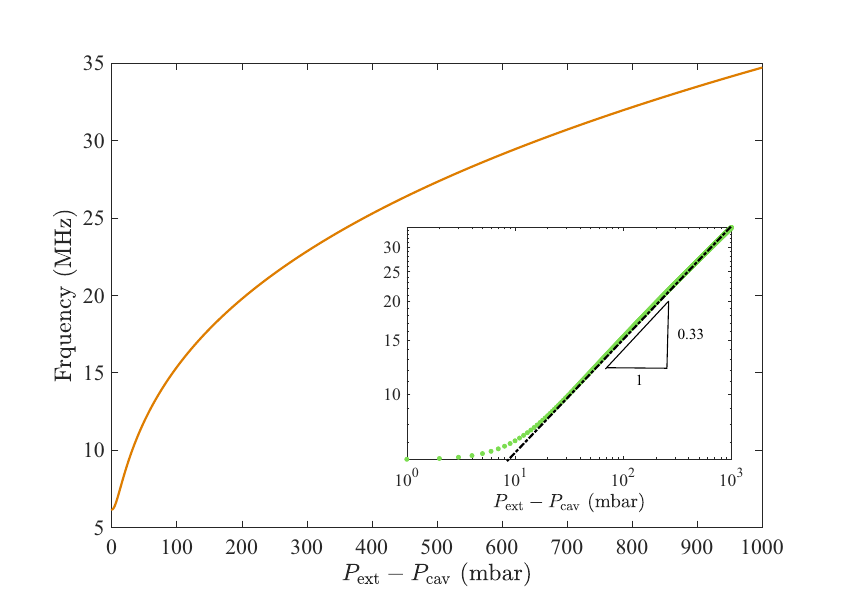}
 \caption{\textbf{Pressure-frequency response obtained from FEM simulations} (see Fig. (3) of the main text). As can be seen, at relatively high pressures, the system shows a 1/3 nonlinearity ($\Delta p \propto {f^{{1 \mathord{\left/
 {\vphantom {1 3}} \right.
 \kern-\nulldelimiterspace} 3}}}$).}
 \label{fig:NonlinearityOrder}
\end{figure}
 
It is worth noting that Eq. (\ref{eq:GenEq1}) only retains the cubic term in the high pressure regime, hence $\Delta p \propto {\left( {\tilde q_1^s} \right)^3}$. By applying the same reasoning, it is possible to demonstrate that Eq. (\ref{eq:GenEq2}) results in $f \propto \left( {\tilde q_1^s} \right)$ and thus $f \propto \Delta {p^{{1 \mathord{\left/
 {\vphantom {1 3}} \right.
 \kern-\nulldelimiterspace} 3}}}$. This trend may be seen in Fig. (\ref{fig:NonlinearityOrder}) as we get closer to higher pressure regimes. 
 
We also note that the model is obtained in modal coordinates. Nonetheless, the governing partial differential equations (Eq.~\eqref{eq:eqNLPlaqUWAxi}) were formulated with in-plane ($u$) and out-of-plane ($w$) displacement fields. To retrieve $u$ from the modal equations, one must use Eqs.~\eqref{eq:AdimPlak},~\eqref{eq:DecomposBMMembrB}, and~\eqref{eq:ContInPlane} simultaneously as follows:
\begin{equation}
u(r,t) = \frac{{{h^2}}}{R}\sum\limits_{p = 1}^{{N_u}} {\sum\limits_{i = 1}^{{N_w}} {\sum\limits_{j = 1}^{{N_w}} {{\Psi _p}\left( r \right)\frac{{b_{ij}^p}}{{\gamma _p^2}}q_i^sq_j^s} } }.  \label{eq:Inplane}
\end{equation}
However, for the out-of-plane deflection, one should note that separation of static and dynamic modal coordinates results in the separation of static and dynamic transverse displacements. Using Eqs. (\ref{eq:CondensCoeffPoutre}), (\ref{eq:DecomposBMMembrA}) and (\ref{eq:AdimPlak}), we will have:
\begin{equation}
w = {w^s} + {w^d}, \label{eq:CondensCoeffPoutreW}
\end{equation}
where:
\begin{subequations}
\label{eq:W}
\begin{align}
{w^s} = h \sum\limits_{k = 1}^{{N_w}} {{\Phi _k}\left( r \right)q_k^s} &,\label{eq:Ws} \\
{w^d} = h \sum\limits_{k = 1}^{{N_w}} {{\Phi _k}\left( r \right)q_k^d}&.     \label{eq:Wd} 
\end{align}
\end{subequations}
\newcommand{\ra}[1]{\renewcommand{\arraystretch}{#1}}
\setlength{\tabcolsep}{3pt}

\begin{table}[b]
\centering
\ra{1.2}
\caption{Benchmarking the present model against the analytic formulations available in the literature}
\begin{tabular}{lccc}
    \toprule
Model  & A & B & C \\
\midrule
Ref.~\cite{Bunch2008} & 4 & \(\displaystyle \frac{8}{{3(1 - {\nu})}}\) & \(\displaystyle \frac{2}{{3\left( {1 - \nu } \right)}}\)          \\
Ref.~\cite{Lee2014} & 4 & \(\displaystyle \frac{8}{{3(1.026 - 0.793\nu - 0.233{\nu^2})}}\) & \(\displaystyle \frac{2}{{3\left( {1 - {\nu}^2} \right)}}\)          \\
Current Study & 3.61 & \(\displaystyle 0.737\varpi (\nu )\) & \(\displaystyle {0.612\varpi (\nu )}\)     \\
    \bottomrule
\end{tabular}
\label{table:parameters}
\end{table}

\section{Comparison of pressure-frequency response with the force-deflection response of AFM in terms of ideal forcing regimes for extracting Young's modulus}
\label{Sec:StressDistribution}
In this section, we use FEM simulations to compare the loading required to obtain Young's modulus using AFM and our proposed technique. To this end, we perform simulations for a pre-tensioned membrane ($n_0=\,\SI{0.3}{\newton\meter^{-1}}$) with a radius of $\SI{5}{\mu\meter}$ and a thickness of $\SI{20}{n\meter}$, composed of polysilicon with Young's modulus of $\SI{160}{GPa}$, density of $\SI{2330}{kg\meter^{-3}}$, and Poisson's ratio of 0.22. It is important to note that the indentation simulation is carried out assuming a cylindrical indenter with a tip radius of $R_{tip}=\SI{8}{n\meter}$. Furthermore, we expect that the membrane will adhere to the indentor during the indentation and that there will be no slippage between the two. 

By indenting a pre-tensioned membrane with a sharp tip, a large force is applied to a small area, resulting in substantial stress in the contact region. As established in the literature, this problem does not have an analytical solution over the entire range of indentations; instead, it converges to a linear response at small indentations and a cubic response at large indentations~\cite{Komaragiri2005, Wan2003, Vella2017}. However, there is a plausible estimate based on linear and cubic terms for predicting Young's modulus and pre-tension as follows~\cite{CastellanosGomez2012}:
\small
\begin{equation}
\label{eq:indentation}
F = \left[ {\frac{{4\pi E}}{{3\left( {1 - {\nu ^2}} \right)}}\frac{{{h^3}}}{{{R^2}}} + \pi {n_0}} \right]\delta  + \left[ {\frac{{Eh}}{{{R^2}{{\left( {1.05 - 0.15\nu  - 0.16{\nu ^2}} \right)}^3}}}} \right]{\delta ^3},
\end{equation}
\normalsize
where $\delta$ is the center deflection. This equation holds, only if the experimental data contains the entire range of indentations, culminating in a cubic stiffness. By plotting the $F-\delta$ response in log-log coordinates and evaluating the response's slope, one can determine whether it contains the cubic regime ($F \propto {\delta}^3$).

\begin{figure*}[tb]
 \centering
 \includegraphics[width=1\textwidth]{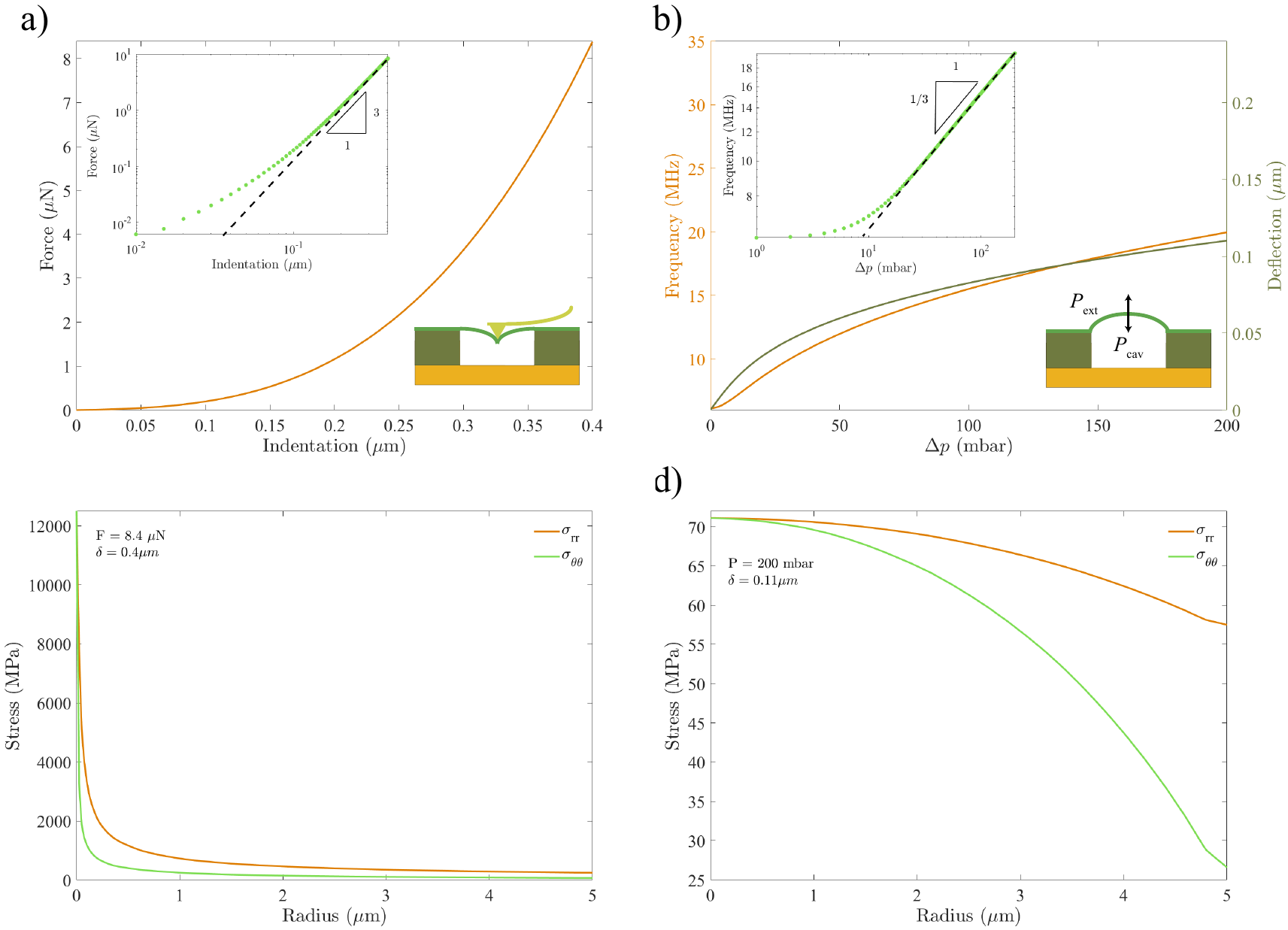}
 \caption{\textbf{Comparison of the minimum loading required for approaching the nonlinear regime for determining Young's modulus in AFM, and pressurization.} a) Force-indentation curve of the center of a membrane under indentation, approaching the cubic stiffness necessary for characterization at $\SI{400}{n\meter}$; b) variations of the fundamental frequency while increasing pressure, reaching one-third stiffness necessary for characterization at $\SI{200}{mbar}$; c) Stress distribution in the membrane showin in subfigure (a), highlighting increased stresses near the contact area in AFM technique; d) stress distribution in the membrane shown in subfigure (b), highlighting more uniformly-distributed stress throughout the membrane.}
 \label{fig:StressDistribution}
\end{figure*}

 
As illustrated in Fig.~\ref{fig:StressDistribution}(a, b), the same membrane approaches the cubic regime at a maximum indentation of 400 nm but reaches the nonlinear pressure-frequency domain ($f \propto \Delta {p^{{1 \mathord{\left/
 {\vphantom {1 3}} \right.
 \kern-\nulldelimiterspace} 3}}}$) at a maximum pressure difference of 200 mbar, resulting in only about 100 nm of center deflection. If we fit these two sets of data using Eq.~(\ref{eq:indentation}) and the pressure-frequency model established in this study, the fitting values are identical; however, the maximum stress created in the indented nanodrum is significantly larger (see Fig.~\ref{fig:StressDistribution}(c, d)). Ref.~\cite{Sharpe2001} collected fracture strength data for polysilicon from several publications and determined that, depending on the size and testing technique, fracture strength values range from 1 to 5 GPa. Based on our basic FEM simulation, the polysilicon drum with the given material and geometrical parameters will fail prior to reaching the required stress regimes for extracting Young's modulus via AFM. As a result, utilizing the pressure-frequency response for extracting Young's modulus of brittle nanomaterials such as polysilicon can be more practical than AFM.
\section{Influence of cavity pressure change and the squeeze film effect on the predicted values of Young's modulus and pre-tension}
\label{Sec:CavityPressChange}
\begin{figure*}[tbp]
 \centering
 \includegraphics[width=0.95\textwidth]{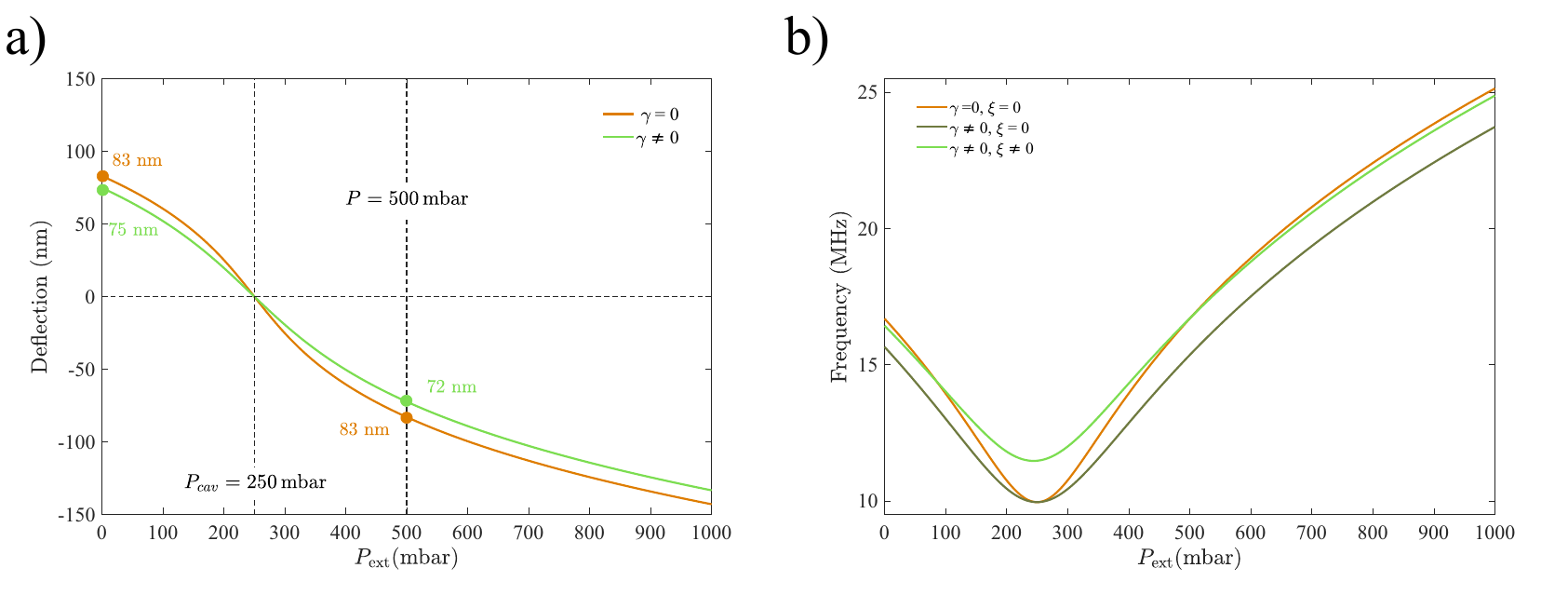}
 \caption{\textbf{Influence of cavity pressure change and squeeze film effect.} a) Membrane deflection with ($\gamma \neq 0$) and without ($\gamma = 0$) cavity pressure change, which exhibits a slight asymmetry around the flat configuration when pressure variations are considered; b) the pressure-frequency response of a membrane considering cavity pressure change ($\gamma \neq 0$) and squeeze film effect ($\xi \neq 0$). The simulations are performed for a pre-tensioned ($n_0=\SI{1.1}{\newton\meter^{-1}}$) with a radius of $\SI{5}{\mu\meter}$ and a thickness of $\SI{33}{n\meter}$, composed of polysilicon with Young's modulus of $\SI{148}{GPa}$, density of $\SI{2330}{kg\meter^{-3}}$, and Poisson's ratio of 0.22.}
 \label{fig:CavityPressChange}
\end{figure*}

As indicated in the main text, deflection of the membrane results in changes in the volume of the cavity beneath it, which in turn yields a change in gas pressure.  The difference between the external and cavity pressure makes the membrane bulge up/down, leading to an increase/decrease of cavity volume and decrease/increase of cavity pressure. Thus, we must include these effects to analyze their importance in determining Young's modulus and pre-tension of the membrane.

Notably, since we are employing single-transverse-mode approximation in this section, the equations can be solved analytically. The equations and parameters used in this section are all in dimensional form.

Using the ideal gas law, we have the following expression:
\begin{equation}
\label{eq:PVequation}
{P_{{\rm{cav}}}}{V_1} = \left( {{P_{{\rm{cav}}}} - \delta p} \right){V_2},
\end{equation}
where ${\delta p}$ is the decrease in pressure due membrane buldging. Rewriting Eq. (\ref{eq:PVequation}) in integral form leads to:
\begin{equation}
\label{eq:PVintegral}
\int_0^{2\pi } {\int_0^R {{P_{{\rm{cav}}}}gr{\rm{d}}r{\rm{d}}\theta } }  = \int_0^{2\pi } {\int_0^R {\left( {{P_{{\rm{cav}}}} - \delta p} \right)\left( {g + w\left( r, t \right)} \right)r{\rm{d}}r{\rm{d}}\theta } } ,
\end{equation}
where $g$ represents the cavity depth, and $w(r, t)$ denotes the deflection of the membrane. Considering axisymmetric deflections, Eq. (\ref{eq:PVintegral}) yields:
\begin{equation}
\label{eq:PVsimplified}
\frac{1}{2}g{R^2}\delta p = \left( {{P_{{\rm{cav}}}} - \delta p} \right)\int_0^R {w\left( r, t \right)r{\rm{d}}r}.
\end{equation}

Employing Bessel-form deflections (Eq. (\ref{eq:ModeShapes})), and setting $N_w=1$ for the sake of simplicity, one can find:
\begin{equation}
\label{eq:deltaP}
\delta p = \frac{{\gamma {\mkern 1mu} {\kern 1pt} {{\tilde q}_1}\left( t \right)}}{{g + \gamma {\mkern 1mu} {\kern 1pt} {{\tilde q}_1}\left( t \right)}}{P_{{\rm{cav}}}},
\end{equation}
where $\gamma=0.4318$. The total pressure difference is the subtraction of external pressure from the cavity pressure:
\begin{equation}
\label{eq:DeltaP}
\Delta p = {P_{{\rm{cav}}}} - \delta p - {P_{{\rm{ext}}}} = \left( {\frac{g}{{g + \gamma {{\tilde q}_1}}}} \right){P_{{\rm{cav}}}} - {P_{{\rm{ext}}}}.
\end{equation}

Inserting Eq. (\ref{eq:DeltaP}) in Eq. (\ref{eq:eqNLPlaqUWAxi}) and following the procedure detailed out in section \ref{Sec:NonlinearCoefs} for ${N_u} \to \infty$ and $N_w=1$ yields:
\begin{equation}
\label{eq:GovPDE}
{\ddot {\tilde {q}}_1} + \bar \omega _1^2{\tilde q_1} + \bar \alpha {\left( {{{\tilde q}_1}} \right)^3} = \bar \beta \left( {\frac{{{g^2}}}{{{{\left( {g + \gamma {{\tilde q}_1}} \right)}^2}}}{P_{{\rm{cav}}}} - {P_{{\rm{ext}}}}} \right),
\end{equation}
where
\begin{equation}
\label{eq:GovPDEparams}
{\bar \omega _1} = \frac{{2.4048}}{R}\sqrt {\frac{{{n_0}}}{{\rho h}}} ,{\mkern 1mu} {\mkern 1mu} {\mkern 1mu} \bar \alpha  = 1.181 \varpi (\nu)\frac{E}{{\rho {R^4}}},{\mkern 1mu} {\mkern 1mu} {\mkern 1mu} \bar \beta  = \frac{{1.6019}}{{\rho h}}.
\end{equation}
We can use Eq. (\ref{eq:CondensCoeffPoutre}) and obtain the following equation for the static bulged configuration:
\begin{equation}
\label{eq:StaticBulgeWithPressureDifference}
\bar \omega _1^2{\tilde q_1^s} + \bar \alpha {\left( {{{\tilde q}_1^s}} \right)^3} = \bar \beta \left( {\frac{{{g^2}}}{{{{\left( {g + \gamma {{\tilde q}_1^s}} \right)}^2}}}{P_{{\rm{cav}}}} - {P_{{\rm{ext}}}}} \right).
\end{equation}
By expanding Eq. \eqref{eq:GovPDE} using Taylor series (up to first order in $\tilde q_1$) and accounting for the squeeze film effect, the equation of motion then becomes~\cite{Dolleman2021}:
\begin{equation}
\label{eq:DynamicBulgeWithPressureDifference}
\ddot{ \tilde{ q}}_1^d + \left[ {\bar \omega _1^2 + 3\bar \alpha {{\left( {\tilde q_1^s} \right)}^2} + \xi } \right]\tilde q_1^d = 0,
\end{equation}
where 
\begin{equation}
\label{eq:squeeze-film}
\xi  = \frac{{2\gamma \bar \beta {g^2}}}{{{{\left( {g + \gamma \tilde q_1^s} \right)}^3}}}{P_{{\rm{cav}}}},
\end{equation}
represents the increased stiffness due to the squeeze film effect.
Thus, adding these two effects to Eq. (\ref{eq:GenEq}),leads to:
\begin{subequations}
\label{eq:GenEquation}
\begin{align}
\left( {\frac{g}{{g + \gamma \tilde q_1^s}}} \right){P_{{\rm{cav}}}} - {P_{{\rm{ext}}}} = 3.61\frac{{{n_0}}}{{{R^2}}}\tilde q_1^s + 0.737\varpi (\nu )\frac{{Eh}}{{{R^4}}}{\left( {\tilde q_1^s} \right)^3},
\label{eq:GenEquation1}\\
f = \frac{{2.4048}}{{2\pi R}}\sqrt {\frac{{{n_0}}}{{\rho h}} + 0.612\varpi (\nu )\frac{E}{{\rho {R^2}}}{{\left( {\tilde q_1^s} \right)}^2} + \frac{{2\gamma \bar \beta {g^2}}}{{{{\left( {g + \gamma \tilde q_1^s} \right)}^3}}}{P_{{\rm{cav}}}}} \label{eq:GenEquation2}
\end{align}
\end{subequations}
where $\tilde{{q}}_1^s$ is the static deflection at the center of the membrane.
By solving Eqs. (\ref{eq:StaticBulgeWithPressureDifference}) and (\ref{eq:DynamicBulgeWithPressureDifference}), we may derive the pressure-frequency response of a pressurized pre-tensioned membrane while accounting for cavity pressure change and squeeze film effect. As can be observed (see Eqs.~(\ref{eq:StaticBulgeWithPressureDifference}) and~(\ref{eq:squeeze-film})), both of these effects are significantly mitigated when the cavity pressure is low, which is also more favorable. Additionally, both phenomena result in asymmetric behavior around the flat configuration, meaning that for a given $\left| {\Delta p} \right|$, the deflection and hence the resonance frequency will be different, as shown in Fig. \ref{fig:CavityPressChange}. However, this asymmetry is negligible. At least for the studied case, Fig. \ref{fig:CavityPressChange} illustrates that combining both phenomena does not significantly modify the response of our experiments at high pressures. This negligible difference will also diminish significantly if we maintain a low cavity pressure. 

In contrast, ignoring squeeze-film effect at low pressures, results in an overestimation of the pre-tension in the membrane. The membrane's bending rigidity, which is not accounted for in our model, is another parameter that can influence the pre-tension estimation for thicker membranes (${h \mathord{\left/
 {\vphantom {h R}} \right.
 \kern-\nulldelimiterspace} R} \ge 0.001$). For instance, the predicted pre-tension value ($n_{0}=1.6~\SI{}{\newton/\metre}$) extracted from the fitting procedure for the sample shown in Fig. 5(a) of the main text, might be inaccurate owing to the aforementioned physical phenomena. The main contributions to this difference could be the added stiffness due to squeeze film effect and the bending rigidity that are not accounted for in the model. If we include the squeeze film effect, pre-tension reduces to $n_0=1.2~\SI{}{\newton/\metre}$ (see Eqs.~\eqref{eq:GenEquation}.  If we also account for the effects of bending rigidity on the pressure-frequency response at the minimum point ($\Delta p = 0$) (see Ref.~\cite{Steeneken2021}), we find a membrane tension value of $n_0=1.15~\SI{}{\newton/\metre}$.

\nocite{*}

\end{document}